\documentstyle[sprocl]{article}

\input{psfig.tex}

\begin{document}

\title{WHAT IS THE PRIMARY SOURCE OF\\
THE MORPHOLOGICAL SEGREGATION?}

\author{S. ANDREON}

\address{Osservatorio di Capodimonte, via Moiariello 16,\\ 80131 Napoli,
Italy\\E-mail: andreon@na.astro.it} 

\maketitle\abstracts{In Coma and two other clusters of galaxies the
primary morphological segregation is the one with respect to a
privileged direction. Only when this segregation is not considered it
appears that the morphological types are segregated in density or in
clustercentric distance.}

\section{Introduction}

In 1880, Wolf first noted the existence of a segregation of
nebulae of different types toward the Virgo cluster. In 1926, Hubble \&
Humason (1926) recognized the general tendency for Es to reside in the core
of clusters.  Dressler (1980) and Dressler et al. (1997) claim that the
morphological types are primarily segregated in galaxy local density,
whereas Whitmore, Gilmore \& Jones (1993) claim that they are segregated
with respect the clustercentric distance. However, as stressed by
Sanrom\'a \& Salvador-Sol\`e (1990) and by Dressler et al. (1997), galaxy
density and clustercentric distance are almost degenerate in galaxy
clusters, except perhaps in irregular clusters, so any correlation with
one of these quantities is likely to be also found with the other one. 

The existence of a observational relation between galaxy properties and
environment imposes the existence of a physical mechanism relating
galaxies and environment. The exact nature of the morphological type
segregation, in density, in clustercentric radius or in whatever
parameter, gives us informations on the nature of the physical mechanism
producing the observed change in the morphological composition of the
clusters.  A clustercentric segregation points out a global mechanism,
whereas a segregation with respect to the local density a local one.

Fig. 1 shows the morphology--density (left panels) and the
morphology--clustercentric distance (right panels) relations. Points mark
our data for a complete sample of $\sim 200$ galaxies in Coma (Andreon et
al. 1996, 1997), lines in the left panels present the morphology--density
relation as determined in 10 nearby centrally concentrated clusters
(Dressler et al. 1997), including Coma. Lines in the right panels show the
morphology--clustercentric distance relation as determined by Whitmore,
Gilmore \& Jones (1993) on 55 nearby clusters (again including Coma).  In
the plot, Dressler density units are transformed in our ones by means of
both estimates for Coma galaxies in common.  Our sample is as deep as
those of Dressler (1980 and 1997) and Whitmore, Gilmore \& Jones (1993)
and it is complete in absolute magnitude, whereas nothing is known on the
completeness of the two comparison samples. Our sample is corrected for
background/foreground interlopers removing galaxies with very high
velocity relative to the cluster center, whereas both comparison samples
are background/foreground statistically corrected because of the scarcity
of redshift measurements for these samples. Our estimate of the
morphological type of Coma galaxies is at least as good as other ones
(Andreon \& Davoust 1997), as indirectly confirmed also by Smail et al.
(1997) and Dressler et al. (1997) which use our morphological types for
confirming the quality of theirs.  Shortly, our sample is smaller,
because composed by just one cluster, but at least as accurate as
the comparison ones.

First of all, we note that our measurements of the morphological
segregation have large errorbars. This is unavoidable for all studies
concerning just a single cluster, because of the limited number of
galaxies in a given cluster. It is for this reason that, in general, the
morphological segregation is studied ``mixing" data from many clusters.

In the studied range of densities of clustercentric distances of Coma, our
points (and likely also all other determinations for any single cluster)
are, given the large errors, compatible with all believable changes of the
morphological composition with radius or clustercentric distances,
including the universal morphology--density and morphology--clustercentric
distance relations (the lines in our graphs) and also with no
morphological segregation at all (horizontal lines in our graphs). A
statistical approach (Andreon 1996) shows that for our sample of Coma
galaxies, the radial and density distributions are statistically
indistinguishable from a morphological types to another one. The same
holds for the Perseus (=Abell 496, Andreon 1994) and Cl0939+4713 (=Abell
851, Andreon, Davoust \& Heim 1997) clusters. Therefore we have no
evidences for a segregation among the morphological types, in these three
clusters on the basis of the above samples. 

A much stronger, statistically significant, morphological segregation, but
of another kind, is instead present {\it in the same sample}. Using the
same Coma galaxies, we are able to put in evidence a segregation with
respect a privileged direction, since many (but not all) morphological
types show an elongated spatial distribution along (or orthogonal to) this
privileged direction (Andreon 1996). Since for a fixed sample we have
detected a relation with respect to a variable (a privileged direction),
but not with respect to another one (the density or clustercentric
distance), then the former relation is stronger that the latter. We found
a similar result for Perseus (Andreon, 1994) and Cl0939+4713 (Andreon,
Davoust \& Heim, 1997). 

\begin{figure}
\centerline{
\psfig{figure=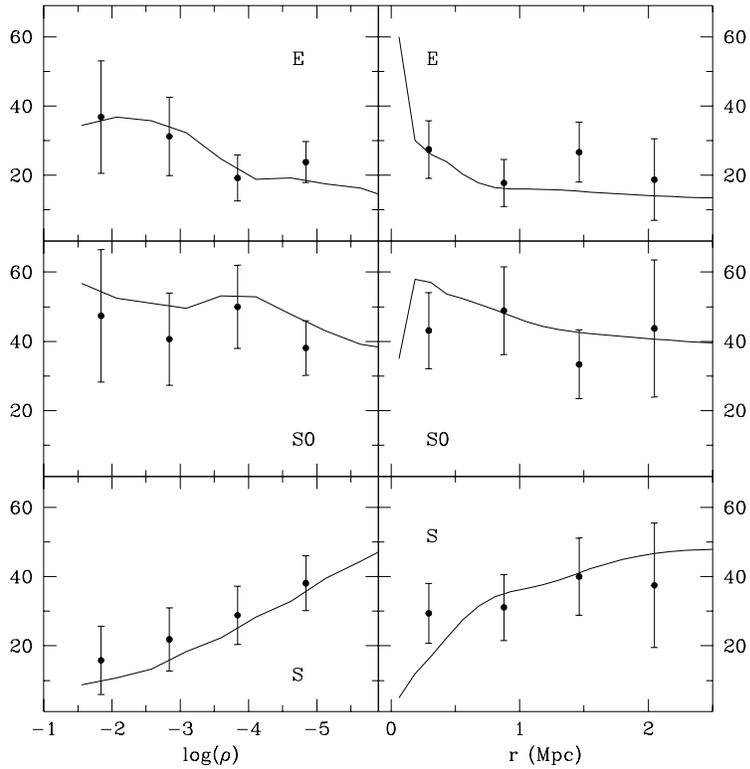,bbllx=30mm,bblly=95mm,bburx=170mm,bbury=240mm,height=10.5truecm}}
\caption{The morphology--density (left panel) and the
morphology--clustercentric distance (right panel) relations. Points are
relative to
our data, whereas lines are the universal relations. See text for details.}
\end{figure}

So, in the three studied clusters, the primary morphological segregation
found is the one with respect to the privileged direction and not in
density or in distance from the cluster center. 

In our three clusters, the privileged direction is roughly aligned with
the major axis of the X-ray images (details are in Andreon, Davoust \& Heim
1997). Here we do not claim that the elongation of the X-ray emission is
the source of a privileged direction in the morphological
segregation, but just that there is an approximate coincidence between
these two directions. Furthermore, the privileged direction in Coma is also
approximatively aligned with the direction of the NGC4839 group, which could 
be, in turn, the reason for the existence of the found privileged directions.

This type of segregation cannot be detected in ``virtual'' clusters formed
by ``mixing'' real ones, as usually done in literature for measuring a
morphological segregation, since the superposition of the clusters destroys
all azimuthal dependences of the segregations as long as all clusters do
not show the same privileged direction in the sky (a pre-copernican
hypothesis). 

To conclude, in Coma and two other clusters of galaxies the primary
morphological segregation is with respect to a privileged direction. Only
when this segregation is not considered it appears that the morphological
types are segregated in density or in clustercentric distance. 

\section*{Acknowledgments}

I thank L. Buson for an attentive lecture of the manuscript, my former PhD
thesis supervisor, E. Davoust and the French Telescope Time Allocation
Committee, that allowed such an expensive observational research to be
carried out.

\end{document}